\documentclass[prc,aps,showpacs,floatfix,nofootinbib,twocolumn,letterpaper,reprint]{revtex4-1}
\usepackage{epsfig} 
\usepackage{amssymb} 
\usepackage{amsmath} 
\usepackage{amsfonts} 
\usepackage{color, graphicx} 
\usepackage{bm}
\newcommand{\ssec}[1]{\emph{#1}.---} 
 
\def\lb{\langle} 
\def\rb{\rangle} 
 
\def\be{\begin{equation}} 
\def\ee{\end{equation}} 

\def\Tr{{\rm Tr}} 
\def\tr{{\rm tr}} 

\def\Im{{\rm Im\,}}
\def\pf{{\rm pf}}

\newcommand{\ket}[1]{\left | #1 \right >} 
\newcommand{\bra}[1]{\left < #1 \right |}

\newcommand{\avg}[1]{\lb #1 \rb}

\begin{document}
\title{Projection after variation in the finite-temperature Hartree-Fock-Bogoliubov approximation}
\author{P. Fanto$^1$}
\affiliation{$^{1}$Center for Theoretical Physics, Sloane Physics
Laboratory, Yale University, New Haven, Connecticut 06520}
\date{\today}
\begin{abstract}
The finite-temperature Hartree-Fock-Bogoliubov (HFB) approximation often breaks symmetries of the underlying many-body Hamiltonian.  Restricting the calculation of the HFB partition function to a subspace with good quantum numbers through projection after variation restores some of the correlations lost in breaking these symmetries, although effects of the broken symmetries such as sharp kinks at phase transitions remain.  However, the most general projection after variation formula in the finite-temperature HFB approximation is limited by a sign ambiguity.  Here, we extend the Pfaffian formula for the many-body traces of HFB density operators introduced by L.~M. Robledo in Ref.~\cite{Robledo2009} to eliminate this sign ambiguity and evaluate the more complicated many-body traces required in projection after variation in the most general HFB case.  We validate our method through a proof-of-principle calculation of the particle-number-projected HFB thermal energy in a simple model. 
\end{abstract}
\maketitle

\ssec{Introduction}The Hartree-Fock-Bogoliubov (HFB) approximation is an important mean-field method for studying many-fermion systems in which pairing correlations are important.  When extended to finite temperature \cite{Goodman1981}, this method provides an efficient way to calculate statistical observables.  The finite-temperature HFB approximation has been widely applied to study the deformation and pairing properties of nuclei \cite{Goodman1986, Egido2000} and is also useful for the study of atomic Fermi gases \cite{Zhao2010}.  However, the finite-temperature HFB approximation often breaks symmetries of the underlying many-body Hamiltonian.  In particular, the HFB approximation explicitly violates particle-number conservation if the pairing field is nonzero and can also violate rotational invariance when the mean-field solution is deformed.  Breaking these symmetries reduces the accuracy of HFB predictions of statistical properties such as nuclear level densities \cite{Alhassid2016}.

To avoid breaking symmetries, the conservation of the symmetries of the underlying many-body Hamiltonian must be enforced during the variation to determine the HFB Hamiltonian.  This procedure is known as variation after projection (VAP) and has been applied successfully in the zero-temperature HFB approximation \cite{Ring1980, Schmid1987}.  However, while at zero temperature the HFB solution is determined by minimizing the energy, at finite temperature the HFB solution is determined by minimizing the grand thermodynamic potential \cite{Blaizot1986}.  Determining this potential in VAP requires calculating the symmetry-projected entropy $S_\Lambda = -\Tr \left[\hat{D}_\Lambda \ln \hat{D}_\Lambda \right]$, where $\hat{D}_\Lambda = \hat{P}_\Lambda \hat{D} \hat{P}_\Lambda$
is the projected HFB density operator, $\hat{P}_\Lambda$ is the projection operator onto the quantum numbers $\Lambda$ of the subspace to which the variation is being restricted, $\hat{D}$ is the unprojected HFB density operator, and the trace $\Tr$ is over all many-particle states in Fock space.  Evaluating this expression for the entropy is complicated and, to date, has only been done in the finite-temperature Bardeen-Cooper-Schrieffer (BCS) approximation by computing the trace explicitly in the many-particle subspace of the full Fock space defined by the projection operator $\hat{P}_\Lambda$ \cite{Esebbag1993,Gambacurta201213}.  Such explicit evaluation in either the BCS or HFB approximation is only possible if this many-particle subspace is sufficiently small or if unbroken symmetries render the necessary matrix algebra tractable.  Consequently, the use of VAP for the restoration of, for example, particle-number conservation or rotational symmetry is currently unfeasible for realistic finite-temperature HFB calculations of mid-mass or heavy nuclei because the combinatorial growth of the dimension of the allowed many-particle subspace with the number of single-particle orbitals renders the evaluation of the projected entropy intractable.

Alternatively, the projection may be applied in the calculation of the HFB partition function but not in the variation that determines the HFB Hamiltonian.  This approach is known as projection after variation (PAV) \cite{Rossignoli1993, Rossignoli1994,Tanabe2005}.  Unlike VAP, PAV is tractable for finite-temperature calculations in large systems such as heavy nuclei \cite{Fanto2017}.  However, PAV does not fully eliminate the effects of broken symmetries.  In particular, sharp kinks around phase transitions occur in PAV observables, and the thermodynamic entropy may become unphysically negative in the low-temperature limit \cite{Fanto2017}.  Despite these problems, finite-temperature HFB calculations with PAV include correlations that are missing in the standard HFB approximation and therefore are more accurate than standard finite-temperature HFB calculations \cite{Alhassid2016}.

However, in the most general HFB case, the calculation of the PAV partition function is limited by a sign ambiguity (see Eq.~(3.46) of Ref.~\cite{Rossignoli1994}).  In many physical cases, the HFB Hamiltonian is invariant under an unbroken symmetry that renders the HFB energies at least two-fold degenerate.  For instance, invariance of the HFB Hamiltonian under time-reversal symmetry guarantees two-fold degeneracy of the HFB energies.  This degeneracy is a necessary condition for eliminating the sign ambiguity of PAV via the method of Ref.~\cite{Fanto2017}.  However, there are important physical systems in which no such simplifying symmetries exist and the HFB energies have no degeneracy.  This may occur, for instance, in odd-even and odd-odd nuclei, in which time-reversal symmetry is broken in the HFB.  Similarly, when a cranking term is added to the Hamiltonian, the HFB energies of the rotating frame have no degeneracy \cite{Goodman1974}.  A general formula for calculating the PAV partition function with no sign ambiguity would be useful for these cases.

Here, we introduce a method for calculating the PAV partition function unambiguously in the most general finite-temperature HFB approximation.  Specifically, we extend the Pfaffian formula for the many-body traces of (unprojected) HFB density operators introduced by L.~M. Robledo in Ref.~\cite{Robledo2009} to evaluate the more complicated traces required in the PAV calculation.  The Pfaffian is the square root of the determinant of a skew-symmetric matrix with a well-determined sign (see Appendix A of Ref.~\cite{Robledo2009}).  To demonstrate the validity of our Pfaffian method, we calculate the particle-number PAV thermal energy in a tractable model in which the HFB energies are not degenerate in the paired phase.  The results from our method match the results obtained if the necessary many-body traces for the PAV calculation are evaluated explicitly in the many-particle model space.  The codes and data files necessary to reproduce the results described below are provided in the Supplemental Material repository for this article \cite{supp}.

Recently, Pfaffian formulas have been introduced to overcome a sign ambiguity in the calculation of overlaps of arbitrary HFB states \cite{Robledo2009, Bertsch2012}.  Although the problem of calculating many-body traces that we address here is different from the calculation of these overlaps, our method shares with these methods the idea of using the Pfaffian to circumvent a sign ambiguity.

\ssec{Pfaffian formula for finite-temperature projection after variation}The projection operator onto state $m$ of the irreducible representation $K$ of a symmetry group is formally given by \cite{Bertsch2012}
\be\label{projop}
\hat{P}_{Km} = \frac{d_K}{\Omega_0}\int d\Omega\, \mathcal{R}^{K}_{mm}(\Omega)\hat{R}(\Omega)
\ee
where $d_K$ is the dimension of the representation, $\Omega_0$ is the total volume integral over the group, $\mathcal{R}^{K}_{mm}(\Omega)$ is the diagonal element corresponding to state $m$ of the matrix representation of the group, and $\hat{R}(\Omega)$ is the symmetry operator acting on many-particle states in Fock space.  In the finite-temperature HFB PAV approach, observables at inverse temperature $\beta = 1/T$ can be calculated from the PAV partition function, which is given by
\be\label{proj-part}
\begin{split}
Z_{Km} & = \Tr\left[\hat{P}_{Km} e^{-\beta\left(\hat{H}_{HFB}-\mu\hat{N}\right)}\right] \\
& = \frac{d_K}{\Omega_0}\int d\Omega\,\mathcal{R}^{K}_{mm}(\Omega)\Tr\left[\hat{R}(\Omega)e^{-\beta\left(\hat{H}_{HFB}-\mu\hat{N}\right)}\right]
\end{split}
\ee
where $\hat{H}_{HFB}$ is the HFB Hamiltonian, $\hat{N}$ is the particle-number operator, and $\mu$ is the chemical potential inserted to constrain the average particle number in the grand-canonical ensemble.  The main challenge of PAV is the evaluation of the many-body traces
\be\label{zeta}
\zeta(\beta, \Omega) = \Tr\left[\hat{R}(\Omega)e^{-\beta\left(\hat{H}_{HFB}-\mu\hat{N}\right)}\right]\;.
\ee
We emphasize that the trace in Eq.~(\ref{zeta}) is over the entire many-particle model space.  Our purpose is to show how to evaluate Eq.~(\ref{zeta}) in the most general HFB case, where the HFB energies have no degeneracy.  Throughout this paper, we assume that the model space of the system under investigation consists of a finite number $N_s$ of single-particle orbitals.
 
In Ref.~\cite{Robledo2009}, Robledo derived a Pfaffian formula to evaluate the traces of grand-canonical HFB density operators.  These operators have the form $e^{(1/2) \eta^\dagger \mathcal{R} \eta}$, where the $2N_s \times 2N_s$-dimensional matrix $\mathcal{R}$ has the property that $\sigma \mathcal{R}$ is skew-symmetric, where the matrix $\sigma$ is
\be\label{sigma}
\sigma = \left(\begin{matrix} 0 & 1 \\ 1 & 0\end{matrix}\right)\,,
\ee
and $\eta = \left(a_1,...,a_{N_s}, a_1^\dagger,...,a_{N_s}^\dagger\right)^T$, where $\{a_k,a_k^\dagger\}$ $(k=1,...,N_s)$ are the fermion annihilation and creation operators associated with some model space basis.  Robledo's formula, given in Eqs.~(12,13) of Ref.~\cite{Robledo2009}, is
\be\label{robledo}
\Tr \left[e^{\frac{1}{2}\eta^\dagger \mathcal{R} \eta} \right]= (-)^{\frac{N_s(N_s+1)}{2}} \frac{e^{-\tr\left[\mathcal{Y}\right]/2}}{\det \mathcal{T}_{22}} \pf \left( \begin{matrix}\mathcal{T}_{12}\mathcal{T}_{22}^{-1} & -(1+\mathcal{T}_{22}^T)\\ 1+\mathcal{T}_{22} & \mathcal{T}_{21}\mathcal{T}_{22}^T\end{matrix}\right)
\ee
where the $N_s \times N_s$-dimensional matrices $\mathcal{T}_{ij}$ are the blocks of the $2N_s\times2N_s$-dimensional matrix
\be\label{matrixT}
\mathcal{T} = e^{\mathcal{R}} = \left(\begin{matrix}\mathcal{T}_{11} &\mathcal{T}_{12} \\ \mathcal{T}_{21} & \mathcal{T}_{22}\end{matrix}\right)
\ee
and the exponential term $e^{-\tr\left[\mathcal{Y}\right]/2} = \left(\det \mathcal{T}_{22}\right)^{1/2}$ follows from the Balian-Br\'ezin decomposition \cite{Balian1969} of the operator $e^{(1/2) \eta^\dagger \mathcal{R} \eta}$.  $\pf$ denotes the Pfaffian of a matrix.  As discussed below, for density operators, the sign of $e^{-\tr\left[\mathcal{Y}\right]/2}$ can be determined easily.  We will show that Eq.~(\ref{robledo}) can be used to evaluate the many-body traces $\zeta(\beta,\Omega)$ given in Eq.~(\ref{zeta}) and will determine the sign of $e^{-\tr\left[\mathcal{Y}\right]/2}$ in this more complicated case.

In any particle or quasiparticle basis of the model space, any fermion operator that conserves total particle or quasiparticle number can be written in quadratic form as
\be\label{fermionop}
\hat{A} = \frac{1}{2}\eta^\dagger \mathcal{A} \eta+ A_0
\ee
where $\mathcal{A}$ is a $2N_s \times 2N_s$-dimensional matrix with the property that $\sigma \mathcal{A}$ is skew-symmetric.  A short proof of this result is given in the Supplemental Material \cite{supp}.  Each of the generators $\hat{A}^{(j)}$ of a symmetry broken by the HFB is a fermion operator that conserves total particle number and thus can be written in the form (\ref{fermionop}) in terms of a constant $A_0^{(j)}$ and a $2N_s \times 2N_s$-dimensional matrix $\mathcal{A}^{(j)}$ with the property that $\sigma \mathcal{A}^{(j)}$ is skew-symmetric.    The symmetry operator $\hat{R}(\Omega)$ is expressed in terms of these generators as
\be\label{sym-op}
\hat{R}(\Omega) = \prod_{j}e^{i x_j(\Omega) \hat{A}^{(j)}}
\ee
where the coefficients $x_j(\Omega)$ are $\Omega$-dependent real numbers.  The HFB Hamiltonian conserves total quasiparticle number and therefore can also be written in any particle or quasiparticle basis in the form (\ref{fermionop}) as
\be\label{H-hfb}
\hat{H}_{HFB}-\mu\hat{N} = \frac{1}{2}\eta^\dagger \mathcal{H} \eta + U_0
\ee
where $\sigma\mathcal{H}$ is skew-symmetric and $U_0$ is a constant.
As a concrete example, in the particle basis in which the HFB solution is determined, $\mathcal{H}$ can be expressed as 
\be\label{Hhfb-particle}
\mathcal{H} = \left(\begin{matrix} h-\mu & \Delta \\ -\Delta^* & -(h^T-\mu)\end{matrix}\right) = W\left(\begin{matrix} E & 0 \\ 0 & -E\end{matrix}\right) W^\dagger
\ee
where $h$ is the Hermitian Hartree-Fock potential, $\Delta$ is the skew-symmetric pairing field, $E = \mathrm{diag}(E_1,...,E_{N_s})$ is the diagonal matrix of the HFB quasiparticle energies, and the matrix $W$ is the general Bogoliubov transformation that diagonalizes $\mathcal{H}$ \cite{Ring1980,Blaizot1986}.  The constant $U_0= \tr(h-\mu)/2-\avg{\hat{V}}$, where $\hat{V}$ is the two-body interaction of the underlying many-body Hamiltonian.  The term $\avg{\hat{V}}$ arises from the variation to minimize the grand thermodynamic potential \cite{Blaizot1986}.  Thus, the argument $\hat{R}(\Omega) e^{-\beta\left(\hat{H}_{HFB}-\mu\hat{N}\right)}$ of the many-body trace in Eq.~(\ref{zeta}) is a product of exponentials of operators of the form (\ref{fermionop}).  

Exponentials of fermion operators of the form (\ref{fermionop}) follow the group property \cite{Balian1969}
\be\label{gp1}
e^{\frac{1}{2} \eta^\dagger \mathcal{A}\eta} e^{\frac{1}{2}\eta^\dagger \mathcal{B}\eta} = e^{\frac{1}{2}\eta^\dagger \mathcal{C} \eta}
\ee
where the matrix $\mathcal{C}$ has the property that $\sigma\mathcal{C}$ is skew-symmetric and is determined from the single-particle representation of the group by the matrix equation
\be\label{gp2}
e^{\mathcal{C}} = e^{\mathcal{A}}e^{\mathcal{B}}\;.
\ee 
Consequently, we may rewrite Eq.~(\ref{zeta}) as
\be\label{zeta-trace}
\zeta(\beta,\Omega) = e^{C_0}\Tr\left[e^{\frac{1}{2}\eta^\dagger \mathcal{C}(\beta,\Omega) \eta}\right]
\ee
where the $2N_s\times2N_s$-dimensional matrix $\mathcal{C}(\beta,\Omega)$ is determined by the matrix equation
\be\label{expC}
e^{\mathcal{C}(\beta,\Omega)} =\left(\prod_j e^{ix_j(\Omega) \mathcal{A}^{(j)}}\right) e^{-\beta\mathcal{H}}
\ee
and the constant $C_0 = i\sum_j  x_j(\Omega) A_0^{(j)} - \beta U_0$ depends on the constants $A_0^{(j)}$ and $U_0$ related to the symmetry generators and the HFB Hamiltonian, respectively.  Evaluating the r.h.s.~of Eq.~(\ref{zeta-trace}) using previously developed methods yields the square root of a determinant \cite{Rossignoli1994}.  The undetermined sign of this square root appears for each term $\zeta(\beta,\Omega)$ in the integral over $\Omega$ in Eq.~(\ref{proj-part}) and consequently limits the evaluation of the PAV partition function (\ref{proj-part}).

To overcome this sign ambiguity, we note that the argument of the trace in Eq.~(\ref{zeta-trace}) has the form of a HFB density operator.  Consequently, we may evaluate Eq.~(\ref{zeta-trace}) using Eq.~(\ref{robledo}), where the required matrix $\mathcal{T}$ (\ref{matrixT}) is given by $\mathcal{T} = e^{\mathcal{C}(\beta,\Omega)}$.  As shown below and in \cite{supp}, the matrices $\mathcal{A}^{(j)}$ and constants $A_0^{(j)}$ may be determined from the expressions for the symmetry generators $\hat{A}^{(j)}$ in second quantization.  The matrix $\mathcal{H}$ and constant $U_0$ are outputs of the standard finite-temperature HFB approximation \cite{Goodman1981,Blaizot1986}.  Thus, it is straightforward to calculate $\mathcal{T}$.  

However, to evaluate Eq.~(\ref{zeta-trace}) using Eq.~(\ref{robledo}), we must determine the sign of the factor $e^{-\tr[\mathcal{Y}]/2}$ in Eq.~(\ref{robledo}).  This term is given by the Balian-Br\'ezin decomposition as \cite{Balian1969}
\be\label{etry-prev}
e^{-\tr[\mathcal{Y}]/2} = \bra{\Phi}e^{\frac{1}{2}\eta^\dagger \mathcal{C}(\beta,\Omega) \eta}\ket{\Phi} = \sqrt{\det \mathcal{T}_{22}} e^{i\delta}
\ee
where $\ket{\Phi}$ is the ground state associated with the operators $\{a_k,a_k^\dagger\}$ of the model space basis and $\delta = 0,\pi$ is the undetermined phase factor.  For the density operators considered in Ref.~\cite{Robledo2009}, $e^{-\tr[\mathcal{Y}]/2}$ is real and positive because the expectation value of a density operator in any state is real and positive.  The operator $e^{\frac{1}{2}\eta^\dagger \mathcal{C}(\beta,\Omega) \eta}$ does not have this property, so we must determine $\delta$ directly.  

Using Eq.~(\ref{etry-prev}), we express the phase factor $\delta$ as
\be\label{delta}
\delta = \Im \left[\ln e^{-\tr[\mathcal{Y}]/2}\right] - \frac{1}{2}\Im \left[\ln\left(\det \mathcal{T}_{22}\right)\right]\;.
\ee
In order to determine $\delta$ from Eq.~(\ref{delta}), it is convenient to formulate the problem in a particle basis.  The matrix $\mathcal{T}$ given by Eq.~(\ref{expC}) is obtained by using the matrices $\mathcal{A}^{(j)}$ and $\mathcal{H}$ appropriate to this basis.  In a particle basis, the ground state is the particle vacuum $\ket{0}$.  The symmetry generators conserve particle number, so the symmetry operator leaves the particle vacuum invariant, i.e.~$\hat{R}(\Omega)\ket{0} = \ket{0}$.  Using this fact together with Eqs.~(\ref{fermionop}-\ref{H-hfb}) and (\ref{gp1}), we rewrite Eq.~(\ref{etry-prev}) as
\be\label{etry-eval}
\begin{split}
e^{-\tr[\mathcal{Y}]/2} &= \bra{0}\left[\prod_j e^{\frac{ix_j(\Omega)}{2}\eta^\dagger \mathcal{A}^{(j)}\eta}\right] e^{\frac{-\beta}{2}\eta^\dagger \mathcal{H}\eta} \ket{0} \\ &
= e^{\beta U_0-i\sum_j x_j(\Omega)A_0^{(j)}}\bra{0}\hat{R}(\Omega)e^{-\beta \left(\hat{H}_{HFB}-\mu\hat{N}\right)}\ket{0} \\
& = e^{\beta U_0-i\sum_j x_j(\Omega)A_0^{(j)}}\bra{0}e^{-\beta\left(\hat{H}_{HFB}-\mu\hat{N}\right)}\ket{0}
\end{split}
\ee
The operator in the expectation value on the r.h.s.~of Eq.~(\ref{etry-eval}) is the (unnormalized) unprojected HFB density operator.  Thus, this expectation value is real and positive, and the phase of $e^{-\tr[\mathcal{Y}]/2}$ is set by the complex coefficient $e^{\beta U_0-i\sum_j x_j(\Omega)A_0^{(j)}}$.  Given the phase of $e^{-\tr[\mathcal{Y}]/2}$, we obtain $\delta$ from Eq.~(\ref{delta}).  Once $\delta$ has been determined, we use Eq.~(\ref{robledo}) to evaluate $\zeta(\beta,\Omega)$ in Eq.~(\ref{zeta}) unambiguously.  By repeating this procedure for every quadrature point in the integral in Eq.~(\ref{proj-part}), we can calculate the PAV partition function in the most general finite-temperature HFB approximation.

\ssec{Particle-number projection in pairing model with cranking}To demonstrate that the Pfaffian method described above works, we show here the results of its application to a simple model.  Our model consists of one nucleon species in a single $j$-shell, the $f_{7/2}$ shell, which has eight single-particle orbitals.  The nucleons interact through a pure pairing interaction, and the system is rotating with angular velocity $\omega$ around the $z$-axis.  The Hamiltonian in the rotating frame is
\be\label{crankhm}
\hat{H} =  - G\sum_{m,m^\prime >0} a_m^\dagger a_{\bar{m}}^\dagger a_{\bar{m}^\prime} a_{m^\prime} - \omega\hat{J}_z\;.
\ee
The single-particle orbitals are labeled by the magnetic quantum number $m$, and $\bar{m}$ denotes the time-reversed counterpart of $m$.  To obtain the results shown below, we set $G = 1$, varied $\omega/G$, and assumed half-filling, i.e.~four particles in the shell.  Under time reversal, $\hat{J}_z$ changes sign, so the Hamiltonian (\ref{crankhm}) manifestly violates time-reversal symmetry for nonzero $\omega$.  In the paired phase, the HFB energies for nonzero values of $\omega$ are not degenerate and thus the PAV method of Ref.~\cite{Fanto2017} cannot be applied.  We note that the Hamiltonian (\ref{crankhm}) always preserves the product of time-reversal and a rotation $\pi$ about the $x$ or $y$ axis \cite{Goodman1974}.  As shown in Ref.~\cite{Goodman1974}, while these symmetries may be used to simplify the HFB equations, for $\omega\neq 0$ the HFB energies in the rotating frame have no degeneracy.

In a finite dimensional model space, the projection operator onto $N$ particles can be written as a discrete Fourier sum, and the particle-number PAV partition function is given by
\be\label{pnp-part}
Z_N = \frac{e^{-\beta \mu N}}{N_s} \sum_{n = 1}^{N_s} e^{-i\varphi_n N_s} \Tr\left[e^{i\varphi_n \hat{N}}e^{-\beta\left(\hat{H}_{HFB}-\mu\hat{N}\right)}\right]\
\ee
where $\varphi_n = 2\pi n/N_s$ are quadrature angles and the leading exponential term cancels the dependence of $Z_N$ on the chemical potential.  The particle-number operator $\hat{N}$ is the generator of a $U(1)$ symmetry group and can be written in the form (\ref{fermionop}) in any particle basis as
\be\label{number-op}
\hat{N} = \sum_{m} a_m^\dagger a_m = \frac{1}{2}\eta^\dagger  \left(\begin{matrix}1 & 0 \\ 0 & -1\end{matrix}\right)\eta + \frac{N_s}{2}.
\ee
The HFB Hamiltonian at each inverse temperature $\beta$ value may be written in the form (\ref{H-hfb}), where the matrix $\mathcal{H}$ is given by Eq.~(\ref{Hhfb-particle}) and the constant $U_0 = \tr(h-\mu)/2 - \avg{\hat{V}}$, as discussed above.  We calculated $Z_N$ for $N=4$ particles at each $\beta$ value using Eq.~(\ref{robledo}) together with Eq.~(\ref{etry-eval}) to evaluate the many-particle traces in the Fourier sum in Eq.~(\ref{pnp-part}).   From the PAV partition function, we calculated $E_N = -\partial\ln Z_N/\partial\beta$, the particle-number PAV thermal energy in the intrinsic frame of the rotating shell.  The algorithm used to evaluate the Pfaffians was adapted from Ref.~\cite{Wimmer2012}.  We also calculated the same quantity $E_N$ from the PAV partition function (\ref{pnp-part}) obtained by constructing the many-body matrices in the traces on the r.h.s.~of Eq.~(\ref{zeta}) and then evaluating these many-body traces explicitly in the many-particle model space.  We refer to this latter method as ``explicit projection.''

\begin{figure}[h!]
\begin{center}
\includegraphics[width=1.2\columnwidth]{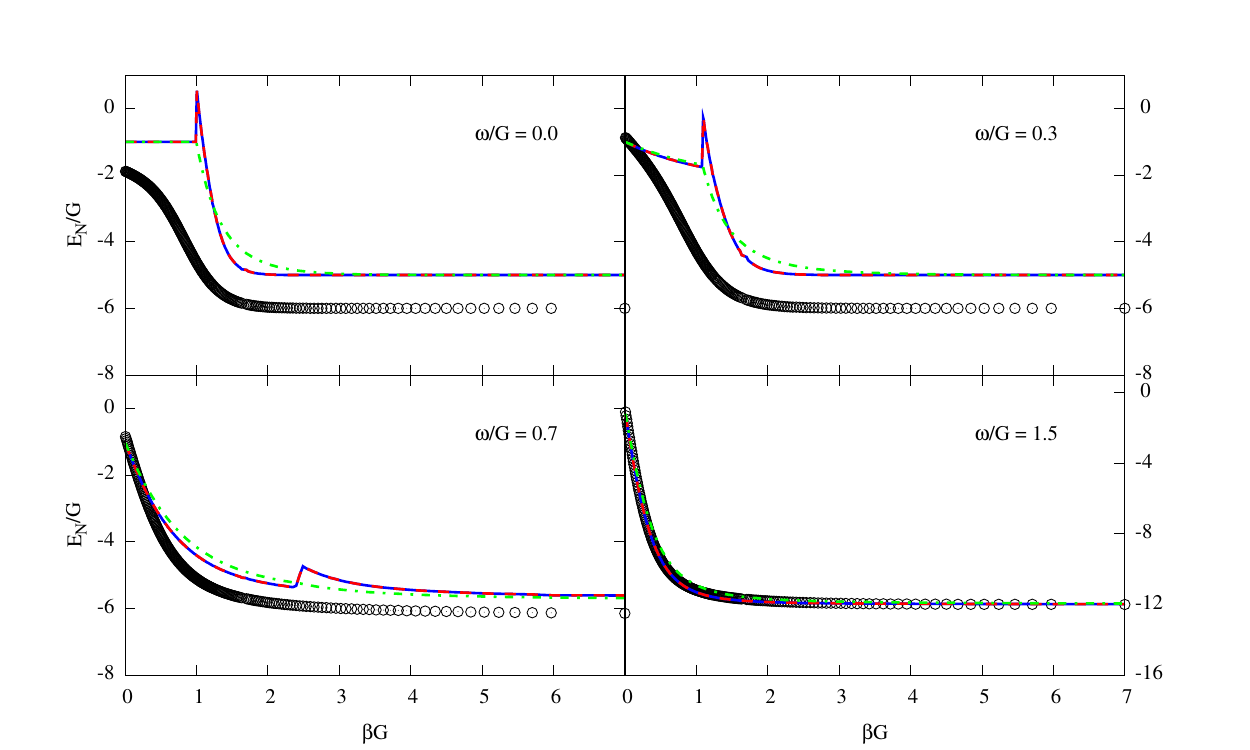}
\caption{\label{energy-comp} The particle-number PAV thermal energy in the rotating frame from the explicit projection (solid blue line) and the Pfaffian formula (red dashed line) as a function of inverse temperature $\beta$.  Open circles show the exact diagonalization results.  The unprojected HFB thermal energy (green dashed-dotted line) is also shown for comparison.  All energies are in units of the pairing strength $G$.}
\end{center}
\end{figure}

In Fig.~\ref{energy-comp}, we compare the PAV thermal energy in the intrinsic frame obtained using the Pfaffian method with the same quantity obtained using explicit projection for a range of $\omega/G$ values.  In every case, the results calculated with the Pfaffian method agree with those from explicit projection to very high accuracy.  This agreement confirms that our PAV method works.  For weak cranking, the PAV energy has a sharp kink at the pairing phase transition.  As $\omega/G$ increases and the single-particle part of the Hamiltonian (\ref{crankhm}) becomes stronger, this kink diminishes in size.  The existence of sharp kinks at phase transitions is common in PAV calculations; e.~g. see Fig.~3 of Ref.~\cite{Fanto2017}.  The kinks in the PAV results in Fig.~\ref{energy-comp} are particularly large for small $\omega/G$ values because the system is very small.  In a VAP calculation, kinks of this type would be smoothed out.

For comparison, we also show the thermal energy in the intrinsic frame from exact diagonalization of the model (\ref{crankhm}) \cite{Bertsch-private} and from the unprojected finite-temperature HFB approximation.  The PAV energy decreases more quickly than the unprojected HFB energy, especially for weak cranking.  The evolution of the HFB results with the cranking term $\omega/G$ is as expected.  With no cranking, the HFB thermal energy is constant in the unpaired phase, and the pairing phase transition is at the expected value $T_c = 1$ \cite{Goodman1974}.    As $\omega/G$ is increased, $T_c$ decreases and the HFB results agree more and more with the exact results.  For $\omega/G = 1.5$, the system is unpaired for all temperatures considered and the agreement between the mean-field results and the exact results is very good.  

For weak and intermediate cranking, comparison of the HFB PAV results with the exact results shows that the PAV method neglects significant correlations.  In particular, the pairing phase transition, which is sharp in the PAV results, is entirely washed out in the exact results.  There is also a significant correlation energy that lowers the exact ground-state energy below the PAV ground-state energy.  In sum, particle-number PAV does not significantly improve over the unprojected HFB results for this simple model.  However, PAV results can be significantly more accurate than results from the unprojected HFB approximation for physically interesting calculations.  We refer the reader to Ref.~\cite{Alhassid2016} for a comparison of approximate particle-number PAV results with unprojected HFB results for heavy nuclei and to Ref.~\cite{Fanto2017} for a benchmarking of particle-number PAV calculations for heavy nuclei against exact results calculated with the shell model Monte Carlo method \cite{Alhassid2016a}.  The calculation done here is intended solely as a proof of the validity of our Pfaffian PAV method.

Finally, since we are proposing that our Pfaffian method be used in realistic calculations, we must consider the scaling of our method's computational time with the model space dimension.  Both the matrix multiplication in Eq.~(\ref{expC}) and the evaluation of the Pfaffian in Eq.~(\ref{robledo}) scale as $O( N_s^3)$ and must be done for each of the quadrature points in the integral in Eq.~(\ref{proj-part}).  In some cases such as the Fourier sum used in the particle-number projection formula (\ref{pnp-part}), there are $N_s$ quadrature points.  Thus, at worst, the overall scaling is $O(N_s^4)$.  For our test case where $N_s = 8$, the time necessary to calculate the PAV partition function and thermal energy for all $238$ $\beta$ values is $\sim 1$ second on a laptop (2.7 GHz Intel Core i5 MacBook Pro with 8 GB of RAM).  The model space used in Ref.~\cite{Fanto2017} for rare-earth nuclei included 40 proton orbitals and 66 neutron orbitals.  The time necessary to run our Pfaffian code for the same number of $\beta$ values in a model spaces of this size would take $\sim1$ hour in the worst case.  Finding the finite-temperature HFB solutions for all the $\beta$ values would consume the bulk of the computational time.

\ssec{Discussion}We have shown how to use the Pfaffian formula for the many-body traces of HFB density operators derived in Ref.~\cite{Robledo2009} to evaluate the more complicated traces necessary to calculate the PAV partition function in the finite-temperature HFB approximation.  We have demonstrated that our Pfaffian method gives correct results by comparing the PAV thermal energy in the intrinsic frame calculated using our method with the same quantity calculated using explicit projection, in which all the many-body traces necessary to calculate the PAV partition function are evaluated directly in the many-particle model space.  It is straightforward to apply our Pfaffian PAV method to any finite-temperature HFB calculation.  The required inputs to our method are (i) the matrices $\mathcal{A}^{(j)}$ and constants $A_0^{(j)}$ defining the generators of the broken symmetry and (ii) the matrix $\mathcal{H}$ and constant $U_0$ defining the HFB Hamiltonian.  $\mathcal{A}^{(j)}$ and $A_0^{(j)}$ may be found analytically from the second-quantized form of the generators in a particle basis, as done in Eq.~(\ref{number-op}).  $\mathcal{H}$ and $U_0$ are outputs of the standard finite-temperature HFB method.  The Pfaffian method developed here allows PAV calculations to be done in the finite-temperature HFB approximation for any system and any broken symmetry.  One interesting application would be finite-temperature HFB studies of odd-mass nuclei with particle-number or angular-momentum PAV.  The techniques developed here could also be used to study the response of the nucleus to rotations at finite temperature by using a cranking model in analogy with the example (\ref{crankhm}) studied above.  

Finally, PAV calculations will always be limited by the effects of the broken symmetries.  VAP completely prevents symmetry-breaking in the HFB approximation, but this method is not yet practical for calculations in large model spaces.  Further developments of VAP methods, such as the use of approximate forms of the entropy as discussed in Ref.~\cite{Tanabe2005}, would be useful.

\ssec{Acknowledgments} 
We gratefully acknowledge Y. Alhassid, G.~F. Bertsch, and L.~M. Robledo for discussions and comments on the manuscript.  We thank G.~F. Bertsch for his help calculating the exact results and verifying the HFB results presented here.  This work was supported in part by the U.S. DOE grant No.~DE-FG02-91ER40608.  Part of this work was completed during the program INT-17-1a, ``Toward Predictive Theories of Nuclear Reactions Across the Isotopic Chart'' at the Institute for Nuclear Theory at the University of Washington.

\end{document}